\documentclass[traditabstract,referee,rnote,longauth]{aa}
\usepackage{amsmath}
\usepackage{txfonts}
\usepackage{graphicx}
\usepackage{natbib}

\newcommand{\hess}{{H.E.S.S.}}
\newcommand{\fer}{{\sl {\it Fermi}}}
\newcommand{\fla}{\fer-LAT}

\newcommand{\gr}{$\gamma$-ray}
\newcommand{\grs}{$\gamma$-rays}

\bibpunct{(}{)}{;}{a}{}{,} 

\begin{document}

\title{The high-energy $\gamma$-ray emission of AP Librae}

\author{H.E.S.S. Collaboration
\and A.~Abramowski \inst{1}
\and F.~Aharonian \inst{2,3,4}
\and F.~Ait Benkhali \inst{2}
\and A.G.~Akhperjanian \inst{5,4}
\and E.~Ang\"uner \inst{6}
\and G.~Anton \inst{7}
\and M.~Backes \inst{8}
\and S.~Balenderan \inst{9}
\and A.~Balzer \inst{10,11}
% ...
}
\institute{
Universit\"at Hamburg, Institut f\"ur Experimentalphysik, Luruper Chaussee 149, D 22761 Hamburg, Germany \and
Max-Planck-Institut f\"ur Kernphysik, P.O. Box 103980, D 69029 Heidelberg, Germany \and
Dublin Institute for Advanced Studies, 31 Fitzwilliam Place, Dublin 2, Ireland \and
National Academy of Sciences of the Republic of Armenia, Yerevan  \and
Yerevan Physics Institute, 2 Alikhanian Brothers St., 375036 Yerevan, Armenia \and
Institut f\"ur Physik, Humboldt-Universit\"at zu Berlin, Newtonstr. 15, D 12489 Berlin, Germany \and
Universit\"at Erlangen-N\"urnberg, Physikalisches Institut, Erwin-Rommel-Str. 1, D 91058 Erlangen, Germany \and
University of Namibia, Department of Physics, Private Bag 13301, Windhoek, Namibia \and
University of Durham, Department of Physics, South Road, Durham DH1 3LE, U.K. \and
DESY, D-15738 Zeuthen, Germany \and
Institut f\"ur Physik und Astronomie, Universit\"at Potsdam,  Karl-Liebknecht-Strasse 24/25, D 14476 Potsdam, Germany
% ...
}

\date{Received ; Accepted}

\abstract{

The $\gamma$-ray spectrum of the low-frequency-peaked BL~Lac (LBL) object
AP~Librae is studied, following the discovery of very-high-energy (VHE;
$E>100\,{\rm GeV}$) $\gamma$-ray emission up to the TeV range by the \hess\
experiment. This makes AP~Librae one of the few VHE emitters of the LBL type. The
measured spectrum yields a flux of $(8.8 \pm 1.5_{\rm stat} \pm 1.8_{\rm
sys}) \times 10^{-12}\ {\rm cm}^{-2} {\rm s}^{-1}$ above 130 GeV and a spectral
index of $\Gamma = 2.65\pm0.19_{\rm stat}\pm0.20_{\rm sys}$. This study also makes
use of \fla\, observations in the high energy (HE, E$>$100~MeV) range, providing
the longest continuous light curve (5 years) ever published on this
source. The source underwent a flaring event between MJD 56306-56376
in the HE range, with a flux increase of a factor 3.5 in the 14-day bin light curve and no significant
variation in spectral shape with respect to the low-flux state. While the
\hess\, and (low state) \fla\ fluxes are in good agreement where
they overlap, a spectral curvature between the steep VHE spectrum and the \fla\, spectrum is
observed. The maximum of the $\gamma$-ray emission in the spectral energy
distribution is located below the GeV energy range.}

\offprints{David Sanchez - email : david.sanchez@lapp.in2p3.fr 
\\ Pascal Fortin - email : pafortin@cfa.harvard.edu
\\ Jonathan Biteau - email : biteau@in2p3.fr
}


\maketitle

\section{Introduction}

The BL Lac class of blazars constitutes about $45\%$ of both the First
(\citealt{2010ApJ...715..429A}; 1LAC) and Second (\citealt{2011ApJ...743..171A};
2LAC) \fer\, Large Area Telescope (LAT) Catalogue of Active Galactic Nuclei
(AGN), and constitutes the majority of the extragalactic very-high-energy (VHE,
E$>$100~GeV) $\gamma$-ray sources\footnote{An up-to-date VHE $\gamma$-ray
catalogue can be found in the TeVCat, {\tt http://tevcat.uchicago.edu}}. AP
Librae falls into the category of ``low-frequency-peaked BL Lac'' (LBL),
defined by an X-ray to radio flux ratio of $f_x/f_r<10^{-11}$
\citep{1995ApJ...444..567P}, and of the more recently introduced ``low frequency
synchrotron peaked'' (LSP) class of blazars defined by a synchrotron emission
peak in the spectral energy distribution (SED) at $\nu_{\rm s,\,peak} \leq
10^{14}\,{\rm Hz}$ \citep[see][]{2010ApJ...716...30A}. This is an order of
magnitude lower than the $\nu_{\rm s,\,peak}$ values found in the bulk of VHE
$\gamma$-ray emitting blazars, which belong to the ``high-frequency-peaked BL
Lac/high frequency synchrotron peaked'' (HBL/HSP) class. A continuity between
these classes of blazars is suggested by the ``blazar sequence''
\citep{1998MNRAS.299..433F}, where the dominance of the high-energy component
and its peak emission energy are inversely proportional to the total luminosity.

AP~Librae was amongst the first few objects to be classified as a member of the
BL Lac class \citep{1972ApJ...175L...7S}, for which a reliable redshift could be
measured ($z=0.049\pm0.002$; \citealt{1974ApJ...194L..79D}). The initial
redshift measurement is consistent with the most recent measurement from the 6dF
galaxy survey ($z=0.0490\pm0.0001$; \citealt{2009MNRAS.399..683J}). An object
coincident with AP~Librae was discovered in the radio band (PKS~1514--24) during
a survey made with the 210-ft reflector at Parkes \citep{1964AuJPh..17..340B},
but it was not until 1971 that the optically variable source AP~Librae and the
radio source PKS~1514--24 were formally associated
\citep{1971ApJ...167L..79B,1971Natur.232..178B}. The host galaxy harbors a black
hole at its center with a mass, estimated using stellar velocity dispersion, of
$10^{8.40\pm0.06}M_\odot$ \citep{2005ApJ...631..762W}.

In X-rays, AP~Librae was first detected by the {\it Einstein X-Ray Observatory}
(1E~1514.7$-$2411; \citealt{1983ApJ...266..459S}). At high energies (HE,
E$>$100~MeV), the source 3EG~J1517$-$2538 \citep{1999ApJS..123...79H} was
tentatively associated with AP~Librae. The photon index reported in the third
EGRET catalog was rather soft ($\Gamma_{\rm HE}=2.66\pm0.43$), resulting in a
low extrapolated flux level in the VHE range covered by atmospheric Cherenkov
telescopes. Observations with the University of Durham Mark 6 \gr\ telescope
resulted in a flux upper limit of $3.7\times10^{-11}\textrm{
cm}^{-2}\textrm{s}^{-1}$ for $E>300\,{\rm GeV}$
\citep{1999ExA.....9...51A,1999ApJ...521..547C}.

An early catalog of bright \gr\ sources detected by the \fla\, was produced
using the first three months of data \citep{2009ApJS..183...46A}. One of these
sources, 0FGL~J1517.9$-$2423, was associated with AP~Librae, but its photon
index was harder \citep[$\Gamma_{\rm HE}=1.94 \pm 0.14$,][]{2009ApJ...700..597A}
than that reported for 3EG~J1517$-$2538. The extrapolation of its spectrum to
higher energies raised the possibility of a detection by Cherenkov telescopes.
In 2010, the \hess\ Collaboration reported the detection of VHE $\gamma$-rays
from AP~Librae \citep{2010ATel.2743....1H}. Following this announcement,
\citet{2010tsra.confE.199F} showed the first radio-to-TeV SED, based on the
preliminary analysis of an HE-VHE data set included in the larger one presented
here, while \citet{2011ICRC....8..199K} also pointed out the existence of an
X-ray jet resolved with Chandra, making Ap Librae the only known TeV BL Lac
object with an extended jet in X-rays.

The paper is organized as follows: in Section \ref{sec:hess}, \hess\
observations are presented while the data analysis of 5 years of \fla\ data is
discussed in Section \ref{sec:fermi}. The variability and broad-band \gr\
emission of AP Librae are discussed in Section \ref{sec:disc}.  

\section{Observations}
\label{sec:observations}

\subsection{H.E.S.S. observations}\label{sec:hess}

\begin{figure}[tbh]
\centering
\resizebox{\hsize}{!}{\includegraphics[width=0.99\textwidth]{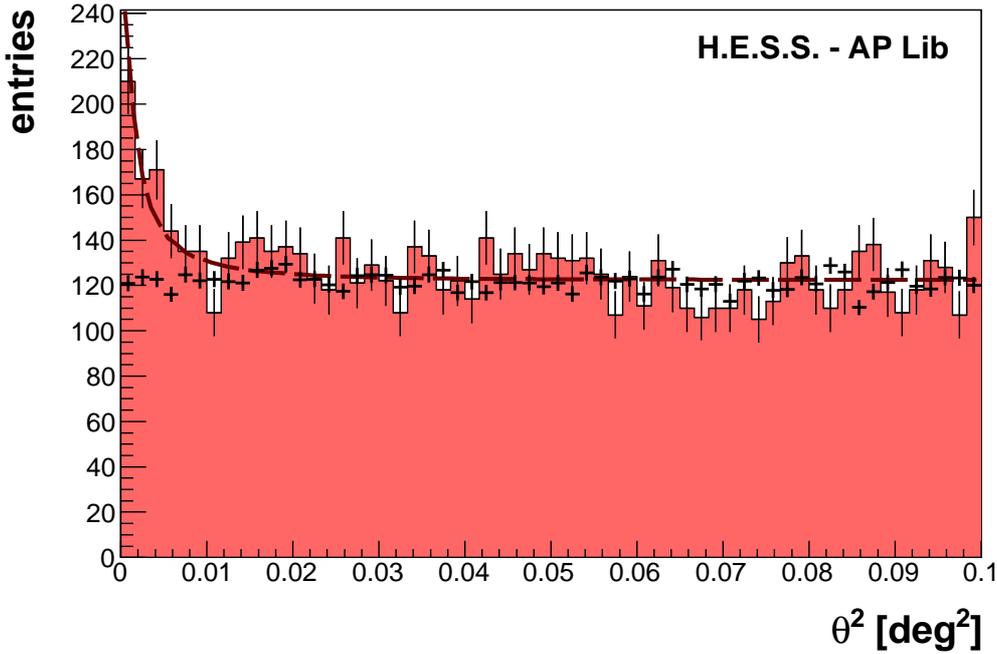}}
\caption{Number of on-source candidate $\gamma$-ray events (solid histogram) and normalized off-source events
  (crosses), as a function of the squared angular distance $\theta^2$ from the position of AP~Librae, compared to a fit of a modeled PSF (dashed line).}
\label{theta2}
\end{figure}

The High Energy Stereoscopic System (\hess), located in the Khomas Highland in
Namibia ($23^\circ16'18''$ S, $16^\circ30'01''$ E), is an array of telescopes
(four at the time of the observations studied) that detect the Cherenkov light
flashes from air showers. \hess\ observed AP~Librae between MJD 55326 (10 May
2010) and MJD 55689 (8 May 2011) for a total of 34 observations of 28 minutes,
each passing data-quality selection criteria (described in
\citealt{2006A&A...457..899A}). This yields an exposure of $14\,{\rm h}$
acceptance-corrected live time with a mean zenith angle of $13^\circ$. In order
to minimize the spectral gap between {\fla} and \hess, cuts achieving the lowest
possible energy threshold were selected. The {\it loose cuts}
\citep{2006A&A...457..899A}, which require a minimum shower image intensity of
40 photoelectrons in each camera, were applied to the data set to perform the
event selection, yielding an average energy threshold of $E_{\rm th}=130\,{\rm
GeV}$. The \textit{Model} analysis method \citep{2009APh....32..231D} was used
to analyze the data within a $0.11^{\circ}$ radius disk centered on the radio
core position of AP Librae \citep[$\alpha_{\rm J2000}= 15^{\rm h}\,17^{\rm
m}\,41.76^{\rm s}$, $\delta_{\rm J2000}=-24^\circ\,22'\,19.6\arcsec$,
][]{1995AJ....110..880J} and further extract the spectrum and light curve, using
the Reflected-Region method \citep{2007A&A...466.1219B} to estimate the
background contamination. With 1133 on-source events, 9042 off-source events and
an on-off normalization of $\alpha=0.10$, the significance of the 218 \grs\,
excess is $6.6\sigma$ \citep[standard deviations, ][]{Lima}. In
Figure~\ref{theta2}, the background (black crosses) and on-source events
distributions (solid histogram) are shown as a function of the squared angular
distance between the source position and the \gr\ direction. The \hess\
point-spread function (PSF) was fitted to the on-source events and matches well
both the signal and the background for large angular distances.

A point-like source model, convolved with the PSF, has been fitted to the data. The
position obtained through this fit is $\alpha_{\rm J2000}=15^{\rm h}\,17^{\rm
m}\,40.6^{\rm s}\pm3.0^{\rm s}_{\rm stat}\pm 1.3^{\rm s}_{\rm sys}$ and
$\delta_{\rm J2000}=-24^\circ\,22'\,37.5''\pm18.4''_{\rm stat}\pm20\arcsec_{\rm
sys}$, compatible within the statistical errors with the location of the AP
Librae core $24''$ away \citep{1995AJ....110..880J}. Further morphological
studies confirm the absence of source extension within the \hess\ PSF.

\begin{figure}[tbh]
\centering
\resizebox{\hsize}{!}{\includegraphics[width=0.99 \textwidth]{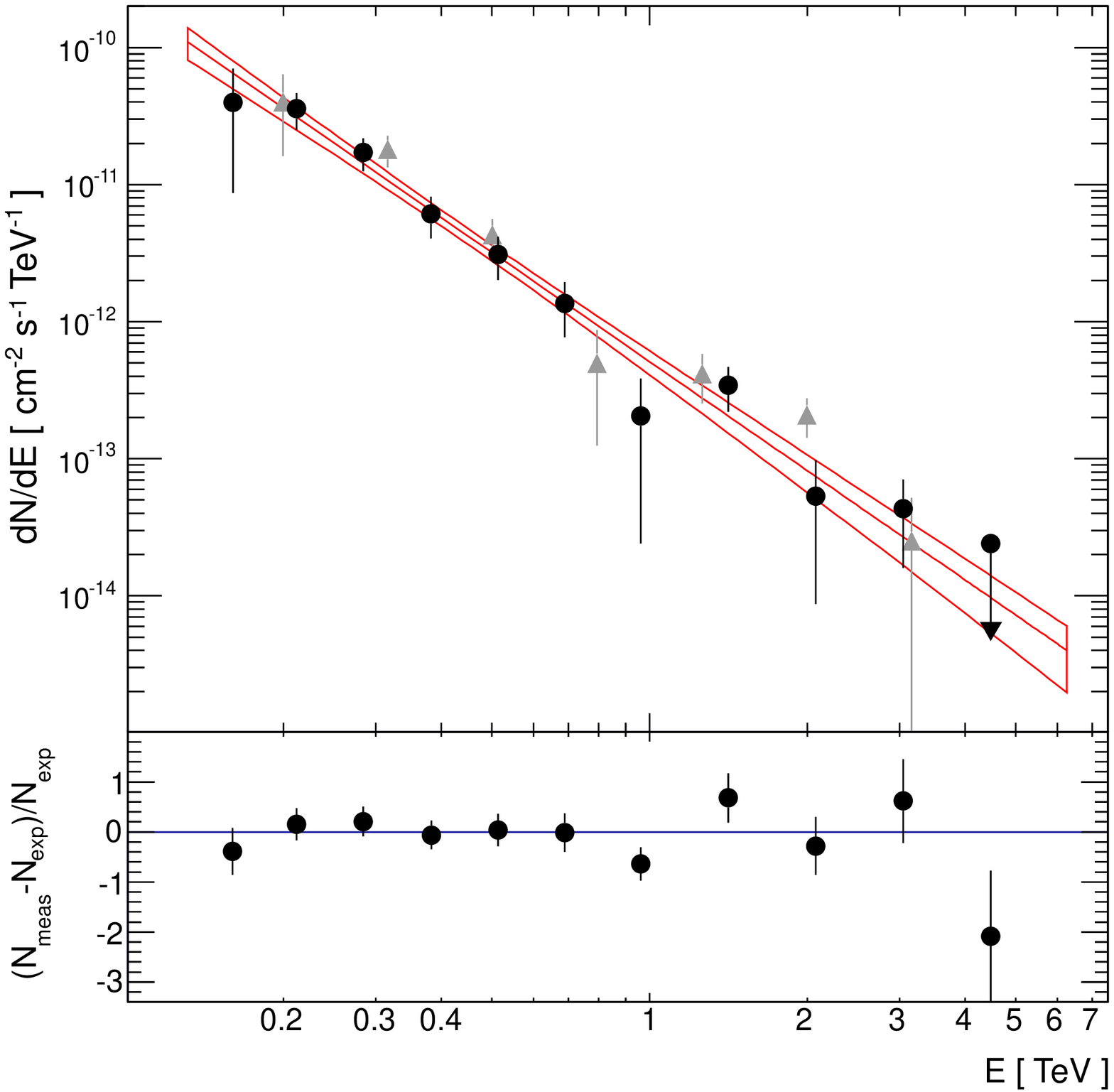}}
\caption{The differential VHE $\gamma$-ray spectrum and corresponding butterfly
of AP~Librae as derived with the \textit{Model} analysis. Uncertainties on the
spectral points are given at $1\sigma$, i.e. at the 68.3\% confidence level, and
upper limits are computed at the 99\% confidence level. The residuals, which are
the difference between the measured and expected number of \grs\, in a bin
divided by the expected number of \grs, are shown in the lower panel. The light
gray points in the upper plot represent the spectrum derived as a cross-check
with the {\it Hillas} analysis. Errors are statistical only.}
\label{fig:spechess}
\end{figure}

The time-averaged photon spectrum for these data is shown in Figure
\ref{fig:spechess}. The best fit is a power-law function, within the energy
range $130\,{\rm GeV}-6.3\,{\rm TeV}$, with a $\chi^2$ probability of
$P(\chi^2)=40\%$, given by: 
\begin{multline} \frac{{\rm d}N}{{\rm d}E}=(4.30
\pm 0.57_{\rm stat}\pm0.86_{\rm sys}) \times 10^{-12} \\ \left(
\frac{E}{E_{\rm dec}}\right)^{-2.65\pm0.19_{\rm stat}\pm0.20_{\rm sys}}
\mathrm{cm}^{-2}\,\mathrm{s}^{-1}\,\mathrm{TeV}^{-1} \label{eq:eq1}
\end{multline}
where $E_{\rm dec}=450\,{\mathrm{GeV}}$ is the decorrelation energy. The
best-fit parameters are obtained using a forward folding technique
\citep{2001A&A...374..895P}. Spectral points are derived with a similar approach
in restricted energy ranges, with a fixed (to the best fit value)
power-law index and a free normalization.

This result was cross-checked with a standard {\it Hillas} analysis
\citep{2006A&A...457..899A} with the {\it loose cuts}, based also on a different
calibration chain. It was found to be entirely compatible with the
\textit{Model} analysis and yielding a detection significance of $6.7\sigma$ and a photon
index of $\Gamma_{\rm VHE}=2.63\pm0.25$ (see also the comparison of both spectra
in Figure \ref{fig:spechess}). The upper limit on the flux derived from
observations taken with the University of Durham Mark 6 \gr\ telescope
\citep{1999ApJ...521..547C}, corresponding to $\sim30\%$ of the Crab Nebula flux
at E$>$300~GeV, is also compatible with the \hess\ spectrum since it is well
above the flux level measured here.

\begin{figure}[tbh]
\centering
\resizebox{\hsize}{!}{\includegraphics[width=0.79\textwidth]{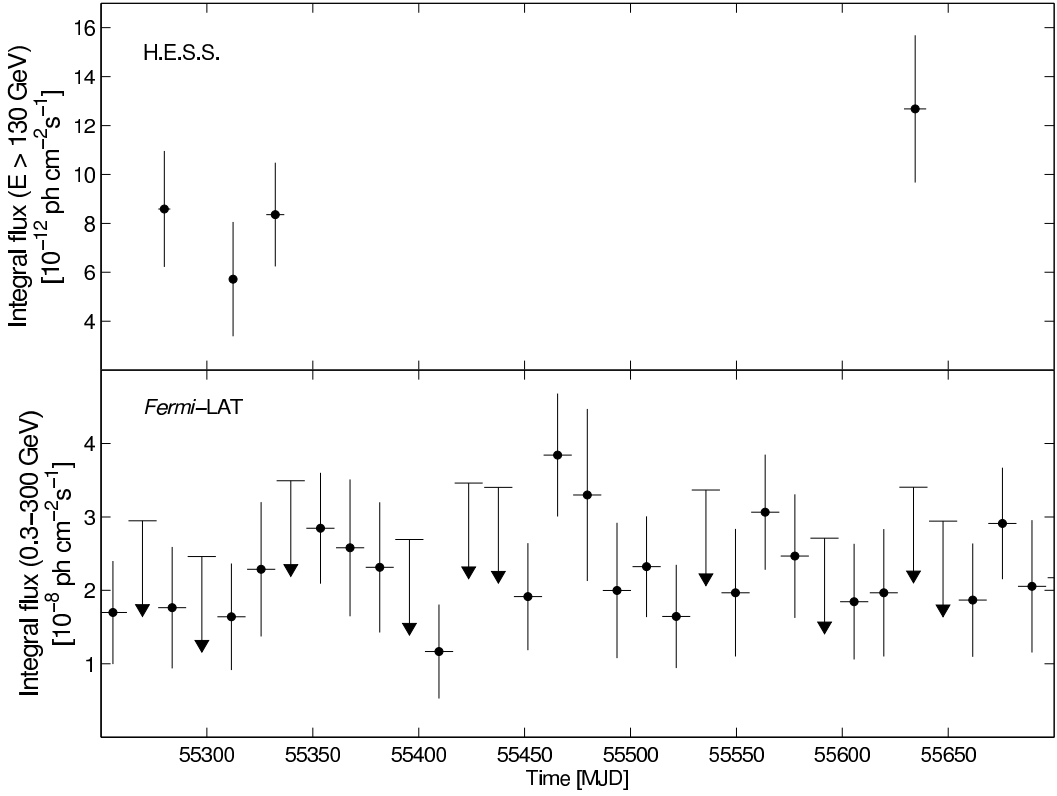}}
\caption{Light curves derived from the observations described in Section
\ref{sec:observations} from MJD~55250 to MJD~55700 (corresponding to the \hess\
measurements). The top panel presents the \hess\ integral flux for $E>130\,{\rm
GeV}$ where the horizontal bars represent the observing duration elapsed between
the two successive full moon periods when \hess\ observed the target. The bottom
panel gives the \fla\, $300\,{\rm MeV}$--$300\,{\rm GeV}$ flux, with $95\%$
confidence level upper limits for segments where ${\rm TS}<10$ and the
horizontal bars show the 14 day \fla\, integration times. This sample is typical
from what was seen throughout the quiescent state.}
\label{FIG:lightcurve}
\end{figure}

The light curve of the integral flux above $130\,{\rm GeV}$, averaged over the
time between two successive full moons, is shown in Figure \ref{FIG:lightcurve}. A
constant function fit to the time series yields a $P(\chi^2)=36\%$ ($\chi^2$/ndf
= 3.2/3), which indicates that the light curve does not show any significant
variability within the observed statistical errors. A 99\% confidence level
upper limit on the fractional variance (as defined in \citealt{Vaughan}) of
$F_{\rm var}<0.46$ is derived \citep{Feldman}. No variability is found using the
{\it Hillas} analysis with the different calibration.

\subsection{\fla\ observations}
\label{sec:fermi}

The \fla, launched on 2008 June 11, is a pair-conversion {\gr} detector
sensitive to photons in the energy range from 20 MeV to more than 300 GeV
\citep{2009ApJ...697.1071A}. The data for this analysis were taken from 4 August
2008 to 4 August 2013 (MJD 54682-56508, 5 years) and were analyzed using the
standard {\fer} analysis software (ScienceTools v9r32p4) available from the
\fer\ Science Support Center (FSSC)\footnote{http://fermi.gsfc.nasa.gov/ssc/}.
Events with energy between 300 MeV and 300 GeV were selected from the {\small
Pass 7} data set. Only events passing the {\small SOURCE} class filter and
located within a square region of side length $20^\circ$ centered on AP~Librae
were selected. Cuts on the zenith angle ($<100^{\circ}$) and rocking angle
($<52^{\circ}$) were also applied to the data. The post-launch
\verb+P7SOURCE_V6+ instrument response functions (IRFs) were used in combination
with the corresponding Galactic and isotropic diffuse emission
models\footnote{http://fermi.gsfc.nasa.gov/ssc/data/access/lat/BackgroundModels.html}.
The model of the region includes the diffuse components and all sources from the
Second \fla\ Catalog \citep[2FGL,][]{2012ApJS..199...31N} located within a
square region of side $24^{\circ}$ centered on AP~Librae. The spectral
parameters of the sources were left free during the fitting procedure. A
power-law correction in energy with free normalization and spectral slope was
applied to the Galactic diffuse component. Events were analyzed using the binned
maximum likelihood method as implemented in {\tt gtlike}.

The source underwent a flaring episode of approximately 10 weeks between MJD
56306-56376 (flaring state). We have therefore defined a quiescent state
measured during the periods MJD 54682-56305 and MJD 56377-56508.

AP~Librae is detected with a high test statistic of TS=2037
\citep[$\approx45\sigma$,][]{1996ApJ...461..396M} in the quiescent state. The
energy spectrum evaluated using this data set is well described by a power-law
with a photon index $\Gamma_{\rm HE}=2.11\pm0.03_{stat}\pm{0.05}_{sys}$, in good
agreement with the 2FGL value, with no significant indication for spectral
curvature. The $300\,{\rm MeV}-300\,{\rm GeV}$ integral flux is
$F_{0.3-300\,{\rm GeV}}=(2.04\pm0.08_{\rm stat}{^{+0.13}_{-0.12}}_{\rm sys})
\times10^{-8}\,{\rm cm^{-2}\,{\rm s^{-1}}}$, and the most energetic photon
within the $95\%$ containment radius of the {\fla} PSF has an energy of 71 GeV.
The systematic uncertainties were evaluated using the bracketing IRFs technique
\citep{2012ApJS..203....4A}.

Replacing the power-law with a log-parabola\footnote{The log-parabola model is
defined as ${\rm d}N/{\rm d}E=N_0 \left(\frac{E}{E_0}\right)^{-(\alpha+\beta \log(E/E_0))}$ }
results in only a marginal improvement in likelihood ($2\Delta\textrm{log}L=10.3$
for 1 degree of freedom, or approximately $3.1\sigma$). With this model,
the best fit differential flux is $N_0= (1.76 \pm 0.10) \times 10^{-13} \,{\rm
ph}\,{\rm MeV^{-1}}\,{\rm cm^{-2}}\,{\rm s^{-1}}$ at $E_0=$5.48\,GeV with an index
$\alpha=2.21\pm 0.06$ and a curvature parameter $\beta=0.07\pm0.02$.

\begin{figure}[tbh]
\centering
\resizebox{\hsize}{!}{\includegraphics[width=0.99 \textwidth]{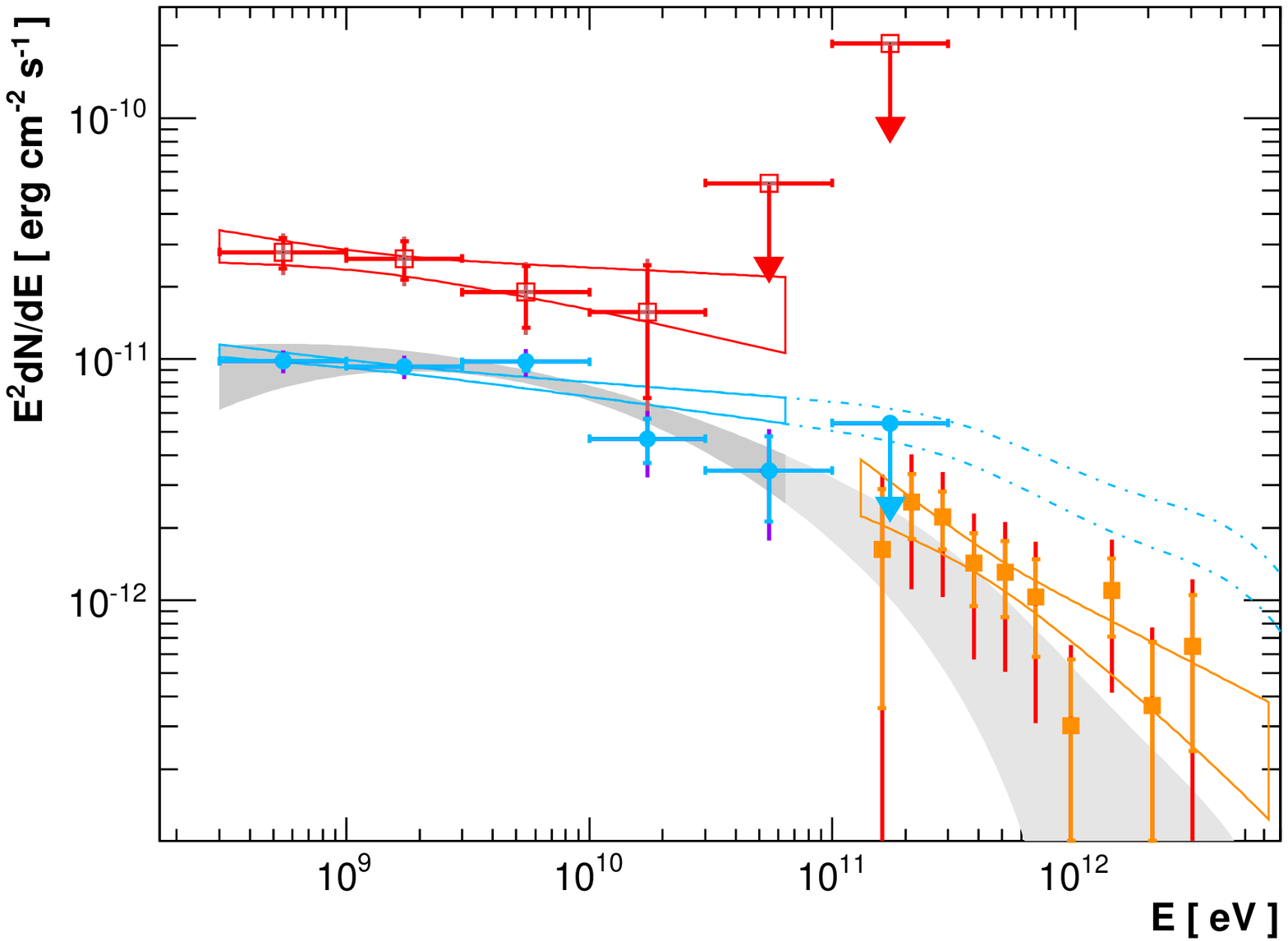}}
\caption{The \gr\ SED of AP~Librae from \fla\, (blue circles) and H.E.S.S.
(orange squares and butterfly power-law fit). For the quiescent state, the
\fla\, best-fit power-law (blue butterfly) has been extrapolated toward the
H.E.S.S. energy range taking EBL absorption into account (dash-dotted line). The
\fla\, log-parabola fit is shown in gray, and its extrapolation taking the EBL
absorption into account is shown in light gray. The flare SED as measured by
\fla\, is given by the red butterfly and open squares. The shorter and longer errors
bars indicate statistical-only and the quadratic sum of statistical and
systematic uncertainties, respectively (cf. text).}
\label{fig:hesed}
\end{figure}

The {\fla} $1\sigma$ spectral error contour for the power-law model of
AP~Librae is presented in Figure \ref{fig:hesed}. Flux values for individual
energy bins were calculated independently, assuming a power-law spectral shape.
For each energy bin, the spectral indices of all sources modeled in the region
of interest were frozen to the best-fit values obtained for the full energy
range and {\tt gtlike} was used to determine the flux. The superimposed
vertical error bars show the statistical uncertainties and the quadratic sum of
statistical and systematic uncertainties, respectively. The latter were
estimated by \citet{2012ApJS..203....4A} to be 10\% of the effective area at
100~MeV, 5\% at 560~MeV and 10\% at 10~GeV and above. 95\% confidence level
upper limits were calculated for energy bins with TS values below 10. For
completeness, the result of the log-parabola fit is also shown in
Figure~\ref{fig:hesed}.

The variability analysis of the LAT data showed a significant flare starting in
2013 January. During the flaring period MJD 56306-56376, the spectrum is well
fitted by a power law with a total flux $F_{0.3-300\,{\rm
GeV}}=(5.55\pm0.57_{\rm stat}{^{+0.36}_{-0.32}}_{\rm sys}) \times10^{-8}\,{\rm
cm^{-2}\,{\rm s^{-1}}}$ and a spectral index $\Gamma_{\rm
HE}=2.11\pm0.09_{stat}\pm{0.05}_{sys}$, consistent in shape with the spectrum
during the quiescent period (see Figure~\ref{fig:hesed}). The peak flux in the two-week bin light curve is $(7.0\pm1.0)\times10^{-8}\,{\rm
cm^{-2}\,{\rm s^{-1}}}$. The flaring state is
discussed more extensively in Section~\ref{subsec:flare}. During this period,
some observations were performed with the H.E.S.S. array, but the resulting data
were too limited to be useful\footnote{Less than 1 hour of useful time was
recorded during the flare. The limited duration and poor background estimation
do not even give a useful limit on the flux.}.

\section{Discussion}\label{sec:disc}
\subsection{The flaring state of AP Librae}\label{subsec:flare}

\begin{figure}[tbh]
\centering
\resizebox{\hsize}{!}{\includegraphics[width=0.99 \textwidth]{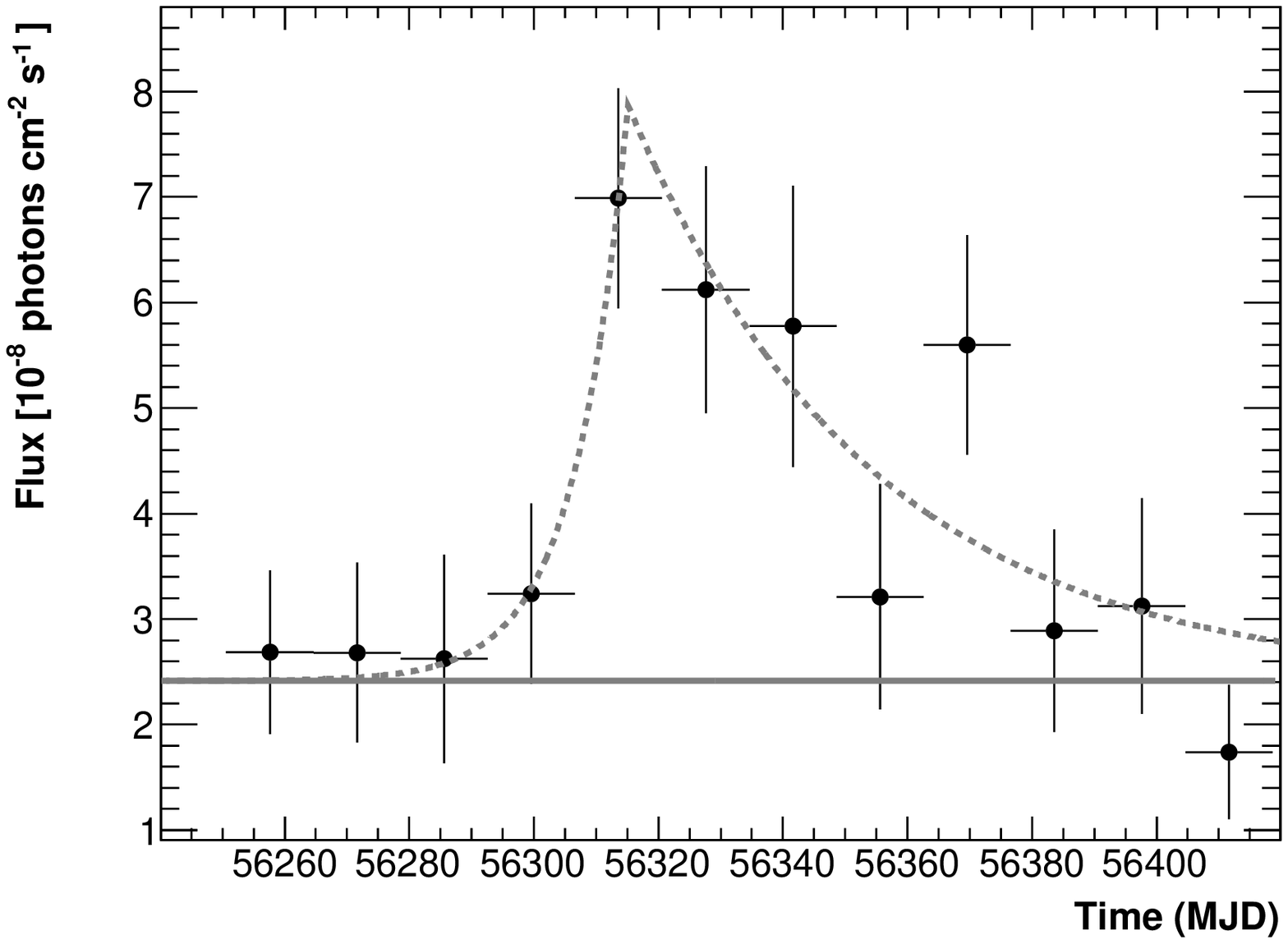}}
\caption{Flux above 300~MeV of Ap Librae during the flare detected by the \fla\ with 14 days integration time. The dashed gray line is the result of the fit with an asymmetric exponential profile (see text) plus a constant (gray line).}
\label{fig:lczoom}
\end{figure}

The \fla\ light curve of the flaring episode above 300~MeV is shown in
Figure~\ref{fig:lczoom}. The peak flux was 3.5 times greater than the
averaged flux. The fastest doubling time scale \citep[as defined
in][]{1999ApJ...527..719Z}, corresponds to the rising part and has a value of
$19 \pm 11$ days. The lightcurve has also been fitted with an asymmetric
profile\footnote{A symmetric Gaussian profile is rejected at a level of
20$\sigma$ with respect to the function used in this work.} $\phi(t) =
A\exp(-|t-t_{\rm max}|/\sigma_{r,d})+B$ where the time of the peak is $t_{\rm
max}$ and the rise and decay time are $\sigma_{r}$ and $\sigma_{d}$. $B$ is a
constant that is also fitted to the data. Fitting this function to the data
yields a peak at $t_{\rm max}({\rm MJD})=56315.1 \pm 2.7$, of amplitude $A = (5.4
\pm 1.4)\times10^{-8}\ {\rm cm}^{-2} {\rm s}^{-1}$, above a constant value of
$B=(2.4 \pm 0.6)\times10^{-8}\ {\rm cm}^{-2} {\rm s}^{-1}$ compatible with the
low-state flux, for a $\chi^2/d.o.f.=6.7/7$. The rise time and
decay time are found to be $\sigma_{r}=5.9 \pm 5.1$ days and $\sigma_{d}=27 \pm 
12$ days, respectively. The rise time (or the doubling time scale) is
compatible with 0 at a $2\sigma$ level, indicating a fast process but the lack
of statistics prevents a more precise probe of this event by making shorter time
bins. Substructures of the flare possibly present on shorter time scale might
be hidden \citep[see][ in which study complex flare structures were found when probing
smaller time scale]{2013ApJ...766L..11S}. An asymmetry in the rise and decay has
already been seen in the GeV range for PKS~1502+106 \citep{2010ApJ...710..810A}
and in the TeV range during the 2006 flare of PKS~2155$-$304
\citep{2007ApJ...664L..71A}. The opposite behavior, i.e. a smaller decay
timescale, has nevertheless been observed in the TeV range in the radio galaxy
M~87 \citep{2012ApJ...746..151A}. However the time scale of the event reported
in this work is much longer and might be of a different origin (e.g. the onset
and decay of large scale structural changes in the jet or possibly a change in
accretion parameters).

\subsection{The LBL AP Librae}

The first evidence for VHE $\gamma$-rays from an LBL-class blazar was the
detection of BL Lacertae ($z=0.069$) at the $5.1\sigma$ significance level
\citep{2007ApJ...666L..17A} corresponding to a flux $3\%$ that of the Crab
Nebula. Its steep VHE spectrum ($\Gamma_{\rm VHE}=3.6\pm0.5$) did not connect
smoothly with the harder \fla\ spectrum \citep{2009ApJ...707.1310A}
($\Gamma_{\rm HE}=2.43 \pm 0.10$) established after the measurement in VHE, but given
the significant variability of the HE $\gamma$-ray flux of BL Lacertae (see
\citealt{2010ATel.2402....1S,2011ATel.3368....1C,2012ATel.4028....1C} and
follow-up ATels), it is possible that the source was in a high VHE flux state at
the time it was detected. Further evidence for VHE $\gamma$-ray emission from
LBL-type objects was found with the detection of S5~0716+714
\citep{2009ApJ...704L.129A}, a source with a steep VHE spectrum ($\Gamma_{\rm
VHE}=3.5\pm0.5$) and a harder HE spectrum \citep{2011ApJ...743..171A}
($\Gamma_{\rm HE}=2.00 \pm 0.02$). It appears that AP~Librae has the smallest
spectral change in the HE-VHE bands, with $\Delta\Gamma=\Gamma_{\rm
VHE}-\Gamma_{\rm HE} \sim 0.56\pm 0.19_{\rm stat}\pm 0.21_{\rm sys}$. We note
however that according to the current classification of extragalactic VHE
$\gamma $-ray emitters in the TeVCaT, (following the classification in the
2LAC), AP Librae would currently be the only VHE $\gamma$-ray emitter of the LBL
class.

With $\Gamma_{\rm HE}=2.11$, AP Librae has a rather soft HE spectrum among the
population of 2FGL AGN also emitting in the VHE regime, for which the average
photon index is $\langle\Gamma_{\rm HE}\rangle = 1.86\pm 0.26$
\citep{2013A&A...554A..75S}. Only the BL Lac object 1RXS~J101015.9-311909, BL~Lacertae,
W~Comae and S5~0716+714 exhibit a spectral index $\Gamma_{\rm HE}\geq2$
\citep{2012ApJS..199...31N}. The observed peak high energy emission $E_{\rm
peak}$ in the SED of objects with $\Gamma_{\rm HE}<2$ is localized roughly above
10~GeV. This quantity, generally not known before the advent of \fer, is of
paramount importance for emission modeling \citep{1998ApJ...509..608T}. Note
that, \citet{2012A&A...542A..94H} reanalyzed the \fer\ data of
1RXS~J101015.9-311909 above 1~GeV and found a hard index of 1.71, which
constrained $E_{\rm peak}$ to be around 100~GeV.

In the next subsection, the peak of the \gr\ emission of Ap Librae is
quantified jointly using the data from \hess\ and \fla.

\subsection{Broad-band gamma-ray emission of AP Librae}

To further investigate the HE-VHE spectral feature, the \fla\, best fit power-law
spectrum was extrapolated to energies greater than $100\,{\rm GeV}$ and
corrected for the EBL attenuation using the model of \cite{2008A&A...487..837F}.
A $\chi^2$ comparison of this extrapolation with the \hess\ spectrum yields a
 $\chi^2/d.o.f.=49/10$ (probability $P(\chi^2)<10^{-6}$). The \hess\
systematic uncertainties were included by shifting the energy by
10\%\footnote{This value is slightly more conservative than the one derived by
\citet{2010A&A...523A...2M} using HE and VHE Crab Nebula data.}, which yields an
uncertainty of $\sigma({\rm
d}N/{\rm d}E)_{\rm sys}=0.1\Gamma_{\rm VHE}\cdot {\rm
d}N/{\rm d}E$ (see Fig.~\ref{fig:hesed}). The same comparison based on an
extrapolation of the log-parabola spectral hypothesis yields a 
$\chi^2/d.o.f.=8.6/10$ (i.e. $P(\chi^2)=57\%$), which suggests broad band
curvature.

\begin{figure}[tbh]
\centering
\resizebox{\hsize}{!}{\includegraphics[width=0.99 \textwidth]{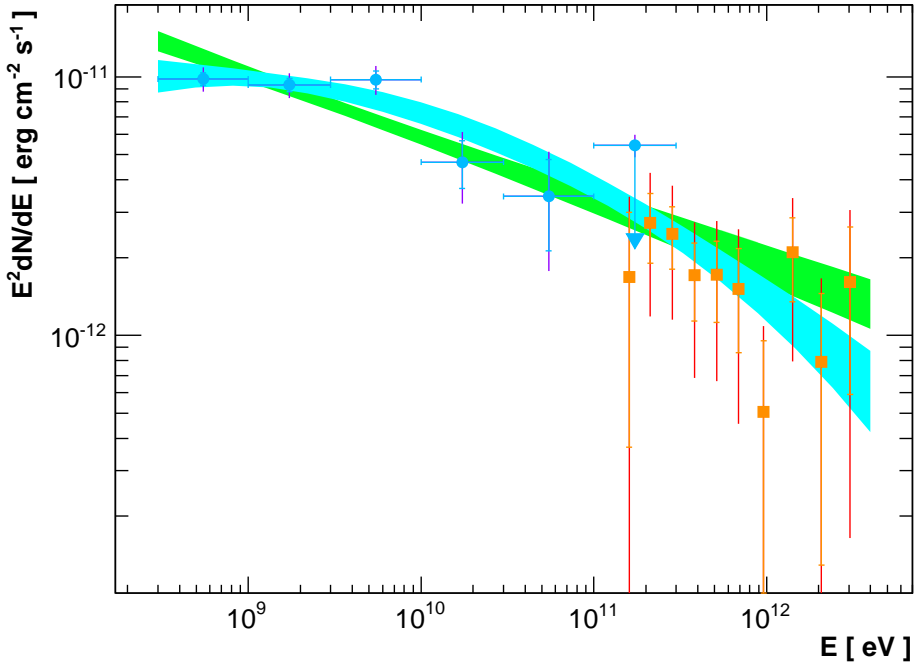}}
\caption{The \gr\ SED of AP~Librae from \fla\, (blue circles) and H.E.S.S.
(orange squares). The green and blue area represent the 68\% error contour of
the power-law and log-parabola fit to the HE-VHE data.}
\label{fig:hesedfit}
\end{figure}

To quantify this curvature, the HE and VHE data points (not corrected for EBL)
were fitted with power-law and log-parabola models, taking into account the
statistical and systematic uncertainties (Fig.~\ref{fig:hesedfit}). In practice,
the fit has been done in log-log space with either a first order (power-law) or
a second order (log-parabola) polynomial function. The parameters obtained are
given in Table \ref{table:PolFit}. The fit of the data with the power-law yields
a $\chi^2/d.o.f.$ of $26.6/13$ (probability of $P(\chi^2) \approx 1$\%), while
the log-parabola yields a $\chi^2/d.o.f.$ of $7.9/12$ (probability of
$P(\chi^2)\approx 79$\%). A likelihood ratio test prefers the latter model at a
level of 4.3$\sigma$, which confirms the presence of curvature in the measured
HE-VHE spectrum of AP Librae. However, the fitting method used for the broad
band HE-VHE data points differs from the methods used within each energy range
and has some limitations (i.e. not taking into account correlations between
energy bins). A proper method to overcome such limitations would consist of a
joint fit of the data, exploiting the response functions of both space-borne and
ground-based $\gamma$-ray instruments, which is beyond the scope of this paper.

\begin{table*}
\caption{Parameters of the first and second degree polynomial functions fit to the HE and VHE data. The functions are of the form $f(x)= log_{10}({\rm d}N/{\rm d}E) = p_{0}+p_{1}x$ and $f(x)= log_{10}({\rm d}N/{\rm d}E) = p_{0}+p_{1}x+p_{2}x^2$ with $x\equiv log_{10}(E/100\,{\rm GeV})$.}
\label{table:PolFit} 
\centering 
\begin{tabular}{c c c c c c} 
\hline\hline 
Model &$ p_{0}$  &  $p_{1}$ & $p_{2}$ &$\chi^2/d.o.f.$\\ 
\hline 
power-law &$ -11.48\pm0.03_{\rm stat}\pm 0.02_{\rm sys}$  &  $-2.25\pm 0.02_{\rm stat}\pm 0.03_{\rm sys}$ & -  & 26.6/13\\ 
log-parabola &$-11.43\pm 0.04_{\rm stat}\pm 0.03_{\rm sys}$  &  $2.37\pm 0.04_{\rm stat}\pm 0.03_{\rm sys}$&$-0.08\pm 0.02_{\rm stat}\pm0.01_{\rm sys}$  & 7.9/12 \\ 

\hline 
\end{tabular}
\end{table*}

Correcting the VHE data points for EBL attenuation and repeating the same joint
fit, the log-parabola model is then preferred at 2.9$\sigma$. In this case the
power-law yields a $\chi^2/d.o.f.$ of $19.4/13$ (probability of $P(\chi^2)
\approx 11$\%) and the log-parabola a $\chi^2/d.o.f.$ of $9.8/12$ (probability
of $P(\chi^2)\approx 63$\%). Scaling up the EBL absorption by thirty percent, as
in \citet{2013A&A...550A...4H}, or using the model of
\citet{2010ApJ...712..238F}, does not significantly affect the latter results,
due to the rather small redshift of the source. 

EBL attenuation is unlikely to be the only explanation of the spectral break
observed in the data. An intrinsic spectral turnover could be due to factors
such as a break in the underlying electron energy distribution, the onset of the
Klein-Nishina regime in the inverse-Compton emission process, or the absorption
of $\gamma$-rays on the circumnuclear radiation fields (see the discussion on
the possibly related phenomenon of GeV breaks observed in the spectra of
flat-spectrum radio quasars: e.g.,
\citealt{2008ApJ...686..181F,2010ApJ...721.1383A,2011ApJ...733...19T,2011ApJ...730L...8A}).
To elucidate this conundrum would require extensive multi-wavelength modeling of
the SED of this complex object\footnote{See, e.g., \citet{2010MNRAS.401.1570T}
who noted the modeling difficulties with simple synchrotron self-Compton
radiative scenarios already when VHE measurements were not yet available and
with a shorter \fla\ exposure than presented here.}, which is beyond the scope
of this Research Note.

The description of the HE-VHE emission of AP Librae by a log-parabola allows
$E_{\rm peak}$ of AP Librae to be estimated at $10^{2.65\pm0.93_{\rm stat}\pm
0.45_{\rm sys}}$~MeV. This value of about 450~MeV is compatible with the
low-energy boundary of the \fla\ range and could then be considered as an upper
limit. It can be compared to the values of $E_{\rm peak}$ determined by
\citet{2010ApJ...716...30A} for the objects BL~Lacertae, W~Comae and
S5~0716+714, using jointly Fermi and publicly available VHE spectra (30~MeV,
4100~MeV and 800~MeV, respectively). Such low-energy emission peaks are rather
uncommon with respect to the bulk of extragalactic VHE emitters, which tend to
have maximum emissions at or above hundreds of GeV. The broad-band emission of
AP~Librae is also rather peculiar, as discussed by \citet{2010tsra.confE.199F}
and \citet{2013ApJ...776...68K}, with an SED dominated by
inverse-Compton and an X-ray spectrum that can not be explained by synchrotron
emission, and that might originate from the same mechanism as the \gr\ emission.
This is consistent with a high-energy component shifted toward lower energies and a peak location that could be below the \fla\ energy range.

\section{Conclusions}

The LBL class of VHE emitting objects proves to be an interesting laboratory to
test radiative model scenarios, and perhaps to identify parameters on which the
LBL-HBL sequence could depend. At present, only a handful of LBL objects have
been detected at VHE (or just this one, depending on the selection criteria),
probably due to a bias toward HBL objects in observation strategies and because
LSP objects are the smallest subset of all $\gamma$-ray selected BL Lac objects
\citep{2013ApJ...764..135S}. Observations with the \hess~II
telescope, and the advent of the Cherenkov Telescope Array (CTA), which will
open the possibility to perform an extragalactic survey (20\% of the sky in 100
hours) with a sensitivity approaching one percent of the flux of the Crab Nebula
\citep{2012arXiv1208.5686D}, should allow more LBL-type blazars to be detected,
and give better insights into the physical processes at work.

\section*{Acknowledgements}

The support of the Namibian authorities and of the University of Namibia in
facilitating the construction and operation of H.E.S.S. is gratefully
acknowledged, as is the support by the German Ministry for Education and
Research (BMBF), the Max Planck Society, the French Ministry for Research, the
CNRS-IN2P3 and the Astroparticle Interdisciplinary Programme of the CNRS, the
U.K. Particle Physics and Astronomy Research Council (PPARC), the IPNP of the
Charles University, the South African Department of Science and Technology and
National Research Foundation, and by the University of Namibia. We appreciate
the excellent work of the technical support staff in Berlin, Durham, Hamburg,
Heidelberg, Palaiseau, Paris, Saclay, and in Namibia in the construction and
operation of the equipment.

The $Fermi$ LAT Collaboration acknowledges support from a number of agencies and
institutes for both development and the operation of the LAT as well as
scientific data analysis. These include NASA and DOE in the United States,
CEA/Irfu and IN2P3/CNRS in France, ASI and INFN in Italy, MEXT, KEK, and JAXA in
Japan, and the K.~A.~Wallenberg Foundation, the Swedish Research Council and the
National Space Board in Sweden. Additional support from INAF in Italy and CNES
in France for science analysis during the operations phase is also gratefully
acknowledged.

The authors want to acknowledge the anonymous referee for his/her help that greatly improved the paper.

\bibliography{aplib_v9_2}
\bibliographystyle{aa}

\end{document}